# Quantum Correlation at Zero-IF: InP HEMT Circuitry Effect

Ahmad Salmanogli
Cankaya University, Engineering Faculty, Electrical and Electronic Department, Ankara, Turkey
salmanogli@cankaya.edu.tr

**Abstract:**
The quantum correlation between microwave modes in an RF electronic circuit is analyzed and studied. An open quantum system operating at 4.2 K is designed in which InP HEMT as the nonlinear component couples two external oscillators to each other. The quantum theory is applied to completely analyze the system, by which the related quantum Hamiltonian containing all noise sources is derived. The Lindblad Master equation is used to analyze the time evolution of the expanded closed system that covers the environmental effects. In the following, the state of the system defined is determined in terms of the ensemble average state using the density matrix; then, the ensemble average of the different operators is calculated. Accordingly, the covariance matrix of the quantum system is derived, and the quantum discord as a key quantity to determine the quantum correlation is calculated. As an interesting point, the results show that InP HEMT mixes two coupling oscillator modes so that the quantum correlation is created at different frequency productions such as $2^{nd}$, $3^{rd}$, and $5^{th}$. The harmonics suitable for sampling and digitalization is the zero-IF (downside of $2^{nd}$ harmonics) band at which the quantum correlation is generated. Another point is that there is no quantum correlation at the frequency resonance of each oscillator coupled to InP HEMT.

**Keywords:** quantum discord, quantum correlation, mixer, InP HEMT, Nonlinearity, covariance matrix

**Introduction:**
One of the critical issues in quantum computing is the interface electronics between quantum processors and digital signal processing units [1-3]. The interface electronics generally include the readout of the qubits and their controlling. The features of the interface circuit can strongly affect the fragile and important quantum properties generated by the quantum processors such as quantum correlation. The latter mentioned property is a fundamental concept in quantum mechanics and can be established between two or more quantum systems [4-8]. It is unlike the classical correlation, which can be understood using conditional probability in classical probability theory. Quantum correlation has been applied in different applications such as quantum information and computation [9-12] and quantum radar [13-15]. One typical example is entanglement, which occurs as two or more quantum systems become strongly correlated. The states of an entangled quantum system cannot be described independently because the state of one system is dependent on the other [16-17]. The other critical quantity that can clarify the quantum correlation and this study focuses on is quantum discord. This quantity measures the quantumness of the correlation between two quantum systems. Quantum discord vanishes just for the classically correlated state, while it is nonzero for all nonclassical states [4-8]. This means quantum discord is more complete than quantum entanglement to cover the quantum correlation features. Consequently, quantum entanglement cannot fully be described over quantum correlation. It is because of the "residual correlation" introduced by the separable mixed states, which any classical probability distributions cannot fit. "For example, in a bipartite system with the separable state as $\rho_{AB} = \sum P_i \, \rho_{Ai} \times \rho_{Bi}$, where $\rho_{Ai}$ and $\rho_{Bi}$ are density matrices of the subsystems, the states $\rho_{Ai}$ and $\rho_{Bi}$ may be physically non-distinguishable. Thus, all information about the subsystems cannot be locally retrieved due to the nonorthogonality of the states [7]". Thus, there are some classically correlated states that may show the signature of quantumness [4-8]. As a result, quantum discord is a more general measure of quantum correlation, which captures all types of quantum correlation.

In this article, we study the time evolution of the quantum correlation (or, more precisely, the time evolution of the quantumness of the correlation) generated because of the InP HEMT (Indium Phosphide High-electron-mobility-transistor) circuitry effects. HEMT is one of the trending transistor technologies that has been applied for very low noise and high frequencies applications [18-19]. The very low noise feature of HEMT, especially InP HEMT, makes it suitable for quantum applications [20-21]. This work tries to demonstrate the quantum correlation generated between microwave modes in a circuit at which two LC oscillators couple to each other through the HEMT. This is the nonlinearity of the HEMT (circuitry effect) that plays the central role. There are some interesting studies in which the quantum correlation between continuous microwave modes has been investigated [22-23]. The mentioned studies focused on the frequency domain and showed the quantum correlation between modes by sweeping the detuning frequency. In contrast, in this work, the time evolution of the quantum correlation is studied. Some interesting features are investigated including the quantum correlation occurring in some specific frequencies due to the mixing of the frequencies. The mixing phenomenon is a nonlinear process created due to the HEMT nonlinearity feature. In a standard electronic circuit as a mixer [24-25], two waves with different frequencies as $f_1$ and $f_2$ are applied to the circuit, and the output frequencies include $f_1 \pm f_2$ as the $2^{nd}$ harmonics. In addition, other harmonics commonly $3^{rd}$ production as $2f_1 \pm f_2$ and $2f_2 \pm f_1$ appear in the output. However, one of the main problems in quantum computing is the dephasing time by which the operation of the quantum system becomes limited [26-27]. Due to the importance of the dephasing time and its impact on the quantum system, this work significantly concentrates on the system's time evolution. It analyzes the quantum discord for the designed system to show how long the quantum correlation between mods can be alive.

**Theory and Background:**
The schematic of the quantum system operating at 4.2 K is illustrated in Fig. 1. The quantum system contains InP HEMT as a nonlinear component, and its internal circuit is shown in the inset figure.

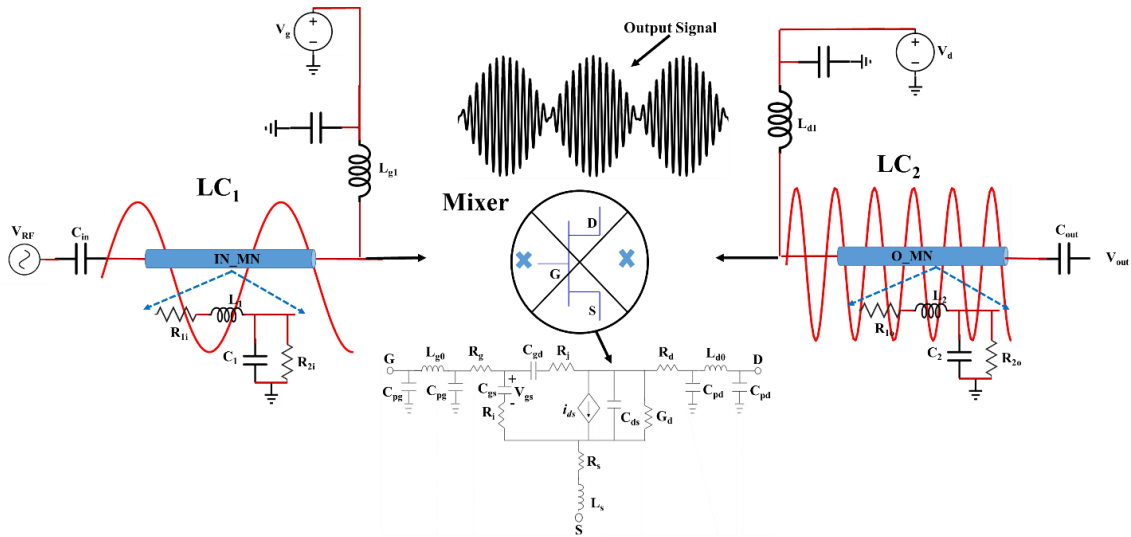

Fig.1 the schematic of the quantum system containing two oscillators coupling through InP HEMT; InP HEMT operates as a mixer shown schematically in the graph.

The mentioned transistor is biased with $V_g$ and $V_d$ through $L_{g1}$ and $L_{d2}$. In addition, the graph schematically shows the input (IN_MN) and output (O_MN) matching network. The circuit is driven capacitively by an RF source ($V_{RF}$). The main goal of this study is to show that InP HEMT nonlinearity mixes the coupled oscillator ($LC_1$ and $LC_2$) waves and generates mixed signals at which the quantum correlation can occur. For this reason, in this figure, the mentioned process is schematically shown in which InP HEMT operates as the mixer, and two input waves with different frequencies ($f_1$ and $f_2$) are mixed. The output signal frequencies differ from the input frequencies. The output frequencies can have 2nd and 3rd frequency productions, such as $f_1 \pm f_2$ and $2f_1 \pm f_2$, respectively. It should be noted that an ideal mixer just generates 2nd frequency.

In the following, we will use quantum theory to analyze the circuit shown and demonstrate the point mentioned. The Hamiltonian of the circuit illustrated in Fig. 1 is theoretically derived, by which the time evolution of the state of the coupled LCs will be investigated. To derive the Hamiltonian, all noise sources effects such as thermal noise generated by the resistors in the circuit and the current source relating noise, are considered [20, 21]. The theory related to the system Lagrangian, classical Hamiltonian and quantum Hamiltonian can be found in [20, 21]. The quantum Hamiltonian of the system ($H = H_0 + H_{int}$), where $H_0$ is the free evolution, and $H_{int}$ is the interaction Hamiltonian, is defined as:

$$H = \{\hbar\Delta_1(a_1^+ a_1) + \hbar\Delta_2(a_2^+ a_2) - \hbar\gamma_{q_1 q_2}(a_1 - a_1^+)(a_2 - a_2^+) - j\hbar\gamma_{q_1 \varphi_2}(a_1 - a_1^+)(a_2 + a_2^+)$$
$$- j\hbar\gamma_{q_2 \varphi_2}(a_2 - a_2^+)(a_2 + a_2^+) - \hbar\gamma_{q_1}(a_1 + a_1^+) + \hbar\gamma_{q_2}(a_2 + a_2^+) - j\hbar\gamma_{q_1}(a_1 - a_1^+) - j\hbar\gamma_{q_2}(a_2 - a_2^+)\}_L \quad (1)$$
$$+ \{-\hbar\gamma_{q_1 q_1 \varphi_2}(a_1 - a_1^+)^2(a_2 + a_2^+) - \hbar\gamma_{q_2 q_2 \varphi_2}(a_2 + a_2^+)(a_2 - a_2^+)^2 + j\hbar\gamma_{q_1 \varphi_2 \varphi_2}(a_1 - a_1^+)(a_2 + a_2^+)^2$$
$$+ \hbar\gamma_{\varphi_2 \varphi_2 \varphi_2}(a_2 + a_2^+)^3 + j\hbar\gamma_{q_2 \varphi_2 \varphi_2}(a_2 - a_2^+)(a_2 + a_2^+)^2 - j\hbar\gamma_{q_1 \varphi_1 \varphi_2}(a_1 - a_1^+)(a_1 + a_1^+)(a_2 + a_2^+)\}_{NL}$$

where $a_i$ and $a_i^+$ (i = 1, 2) stand for the annihilating and creating operator for the defined two oscillators. The first two terms represent the free evolution related to the individual oscillator energy, and other terms determine the interaction Hamiltonian. The interaction Hamiltonian is divided into two parts, linear and nonlinear. All constants in Eq. 1 are defined in Appendix A (Eq. $A_1$). In addition, $\Delta_1$, $\Delta_2$, $\hbar$, and $j$ are the detuning, reduced Planck constant, and the imaginary unit, respectively.

The circuit illustrated in Fig. 1 is considered a multipartite system, so we need to expand the Hilbert space by taking the tensor product for each oscillator component. The dynamics of such a system are governed by the Lindblad master equation [16]. It is necessary to apply the Lindblad master equation rather than the Schrödinger equation because the quantum system interacting with the environment is stochastic. Thus, the state of the open system is described in terms of the ensemble average state. This is done using the density matrix formalism by which the related quantum state's probability distribution is defined. The Lindblad master equation is applied to derive a quantum system's dynamics equation of motion interacting with its environment. In this work, similar to the literature [16, 28], it is supposed that the system is expanded to include the environment; thus, the expanded system (the circuit combined with the environment) can be considered a closed quantum system. A closed quantum system evolution is governed by the von Neumann equation derived as [16, 28]:

$$\dot{\rho}(t) = \frac{1}{j\hbar}[H_t, \rho_t(t)] \quad (2)$$

where $\rho(t)$ is the density matrix describing the probability distribution of a quantum state. $H_t = H_0 + H_{int} + H_{env}$, where the terms, respectively, define resonators free Hamiltonian, interaction Hamiltonian between LC oscillator and InP HEMT, and environmental Hamiltonian. The critical point here is the dynamics of the LC oscillators coupling to the InP HEMT; thus, one can perform the partial trace over $H_{env}$ to attain the master equation. The complete form of the master equation is the Lindblad master equation defined as [28]:

$$\dot{\rho}(t) = \frac{1}{j\hbar}[H, \rho(t)] + \frac{1}{2}\sum_n [2C_n \rho(t) C_n^+ - \rho(t) C_n^+ C_n - C_n^+ C_n \rho(t)] \quad (3)$$

where $C_n$ is the collapse operator by which the quantum system is coupled to the environment. In our latter study [29], we theoretically analyzed a quantum system's coupling to the reservoir and introduced key factors affecting the coupling systems. Finally, the decay rate of the quantum system due to the coupling to the reservoir was derived. However, it should be noted that some approximations are made to derive the master equation in Eq. 3, including 1. There is no correlation between the quantum system and environment at t = 0; 2. The state of the environment doesn't significantly change since interacting with the quantum system; 3. The system and environment remain separable during the evolution; 4. The time scale of the environment is much shorter than the quantum system dynamics. In the following, the Lindblad master equation will be used to govern the time evolution of the system density matrix by which the ensemble average of any operators becomes calculated [28]. In this work, the Qutip toolbox in Python [28] is used to solve Eq. 3. Using the operator's ensemble average resulting from Eq. 3, the time evolution of the covariance matrix (CM) is obtained for the quadrature of the microwave LC oscillators coupling through the InP HEMT nonlinearity. The quadrature of the LC oscillators is defined as $X_i = (a_i + a_i^+)/\sqrt{2}$ and $Y_i = (a_i - a_i^+)/j\sqrt{2}$, i = 1,2.

The compact form of quantum discord (quantum $b$ discord in which the measurement is performed on the second oscillator) is calculated using $D(\rho_{AB}) = h(b) - h(v_-) - h(v_+) + h(\tau + \eta)$, where $v_\pm$ is the Symplectic eigenvalue of the CM and function "h" is defined as $h(x) = (x+0.5)\log_2(x+0.5) - (x-0.5)\log_2(x-0.5)$ [7,13]. In addition, $b$ stands for the second oscillator output photon number, and other quantities $\tau = d_{o12}^2/(b^2-1)$, and $\eta = a - (b \times d_{o12}^2/(b^2-1))$, where $a$ is the first oscillator output photon number, and $d_{o12}$ is the phase-sensitive cross-correlation between two coupled oscillators. The last term, $h(\tau + \eta)$ in the equation, is the effect of the classical correlation depending on the type of measurement performed on the second oscillator [13, 21]. The quantum discord formula indicates that the key factors to increase the quantum discord (quantum correlation between modes) are the second oscillator photon number, the Symplectic eigenvalue of the CM, and finally the classical correlation effect. Nonetheless, it should be noted that the phase-sensitive cross-correlation between two oscillators $d_{o12}$ plays a key role.

Table. 1 Values for the small signal model of the cryogenic InP HEMT [18-19].

|  | **Stands for** | **Value [Unit]** |
| --- | --- | --- |
| $R_g$ | Gate resistance | 0.3 Ω |
| $L_g$ | Gate inductance | 75 pH |
| $L_d$ | Drain inductance | 70 pH |
| $C_{gs}$ | Gate-Source capacitance | 107 fF |
| $C_{ds}$ | Drain-Source capacitance | 51 fF |
| $C_{gd}$ | Gate-Drain capacitance | 60 fF |
| $R_i$ | Gate-Source resistance | 0.07 Ω |
| $R_j$ | Gate-Drain resistance | 8 Ω |
| $G_d$ | Drain-Source conductance | 12 mS |
| $g_m$ | Intrinsic transconductance | 82 mS |
| T | Operational temperature | 4.2 K |
| $T_d$ | Drain conductance noise temperature | 450 K |

**Results and Discussions:**

This section presents the simulation results for the defined quantum system. In all simulations, the data presented in Table.1 is used to model the InP HEMT internal circuit. In addition, it is supposed that the first oscillator state is initiated with the Fock state $|\Psi_1\rangle = |0\rangle$, while the second oscillator is commenced with the coherent state $|\Psi_2\rangle = |\alpha\rangle$. The reason is that we want to study the initial state's effect on the system's time evolution. This study analyzes the step response of the quantum system operating at 4.2 K.

To entirely demonstration of the mixing effects of the InP HEMT, we initially studied a simplified version of the designed system containing only two LC oscillators coupling through a capacitor with the coupling rate $\gamma_c$. In other words, all rates in Eq. 3 except $\Delta_1$, $\Delta_2$, and $\gamma_{q1q2}$ are ignored to reduce the present complex system to a simple LC oscillator coupling through a capacitor. Therefore, the Hamiltonian in Eq. 3 is reduced to $H = \hbar\Delta_1 a_1^+ a_1 + \hbar\Delta_2 a_2^+ a_2 - \hbar\gamma_c(a_1^+ - a_1)(a_2^+ - a_2)$. In this equation, $\gamma_c = 1/C_c(Z_1 Z_2)^{0.5}$, where $C_c$, $Z_1$, and $Z_2$ are the coupling capacitor, and the first and second oscillator impedance, respectively.

For the reduced system and in the following, all the necessary quantities such as the average number of photons, the phase-sensitive cross-correlation between the oscillators, and the quantum discord as a quantifier to present the quantumness are calculated using Eq. 3 in Python (Qutip toolbox). The simulation results for two capacitively coupled LC resonators are shown in Fig. 2. The average number of photons for the second oscillator and quantum discord as a quantifier, and also the related frequency spectrum (FFT: Fast-Fourier Transform) are demonstrated in Fig. 2. Using data listed in Table.1, the oscillators' frequencies are designed to be 5.57 GHz and 5.77 GHz, respectively. The graph in Fig. 2a shows the average number of photons in which there is a small fluctuation around 15.67 and reaches the steady state after ~60 nsec. The average number of photons in the steady state without any small fluctuation relates to the thermal photon number generated in the circuit at 4.2 K. The FFT graph (Fig. 2b) only shows a frequency, which relates to the damping rate of the system. The damping frequency is generated due to the system's interaction with the environment. In addition, the quantum discord and the related FFT are demonstrated in Fig. 2c and Fig. 2d, respectively. The results show that the quantum discord is damped with the same frequency as $N_{ph2}$. As can be seen easily from the graphs, there is no mixing behavior so that no other frequency productions can appear. It should be noted that the behavior of the first oscillator average photon number $N_{ph1}$ is similar to the $N_{ph2}$ depicted in Fig. 2a. For this reason, only the second oscillator photon number is illustrated.

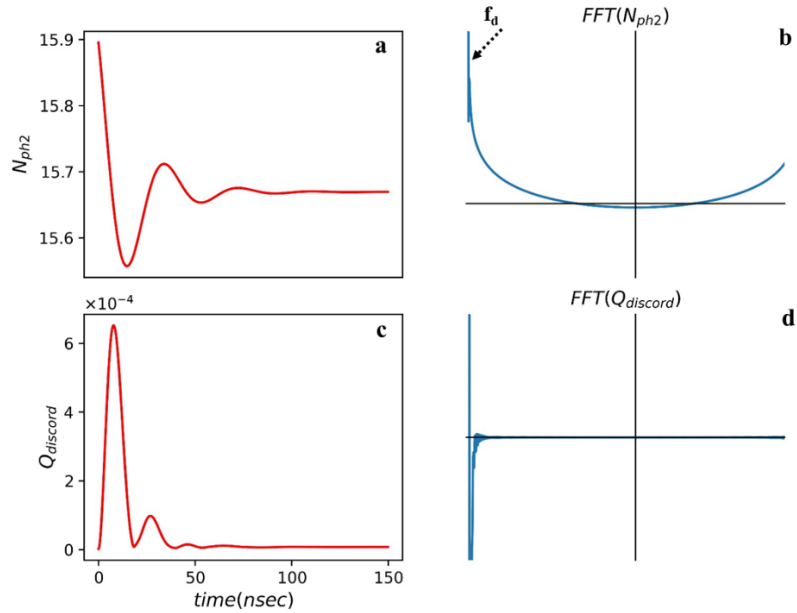

Fig. 2 a) and b) The second LC resonator average photon number and its FFT, c) and d) time evolution of quantum discord and the related FFT; the first and second oscillator relating frequencies are $f_1 = 5.57$ GHz, $f_2 = 5.77$ GHz.

Now, all factors in Eq. 3 are considered, and the simulation results are depicted in Fig. 3. One of the interesting points here in Fig. 3 is the mixing behavior of the quantum system, which is generated due to the nonlinear properties of the InP HEMT internal circuit. From the classical point of view, the designed quantum system is similar to the passive-HEMT mixer circuit [24], in which two sources with different frequencies are applied to the gate and drain of the transistor. The graph for $N_{ph1}$ and its FFT are shown in Fig. 3a and Fig. 3b, respectively. The FFT graph shows three different frequencies; one of them is the damping frequency relating to the quantum system interacting with the environment, and others ($f_1 \pm f_2$), which is the 2$^{nd}$ frequency production, are created due to the mixing behavior. Fig. 3c shows the phase-sensitive cross-correlation ($N_{ph1\&ph2}$) in the designed system. Even though the graph preserves the three mentioned frequencies in Fig. 3b, it also offers some other frequencies related to the nonlinearity effects of the InP HEMT. However, from the quantum mechanical point of view, the arisen frequencies are associated with the off-diagonal elements in the CM and the contributed eigenvalues.

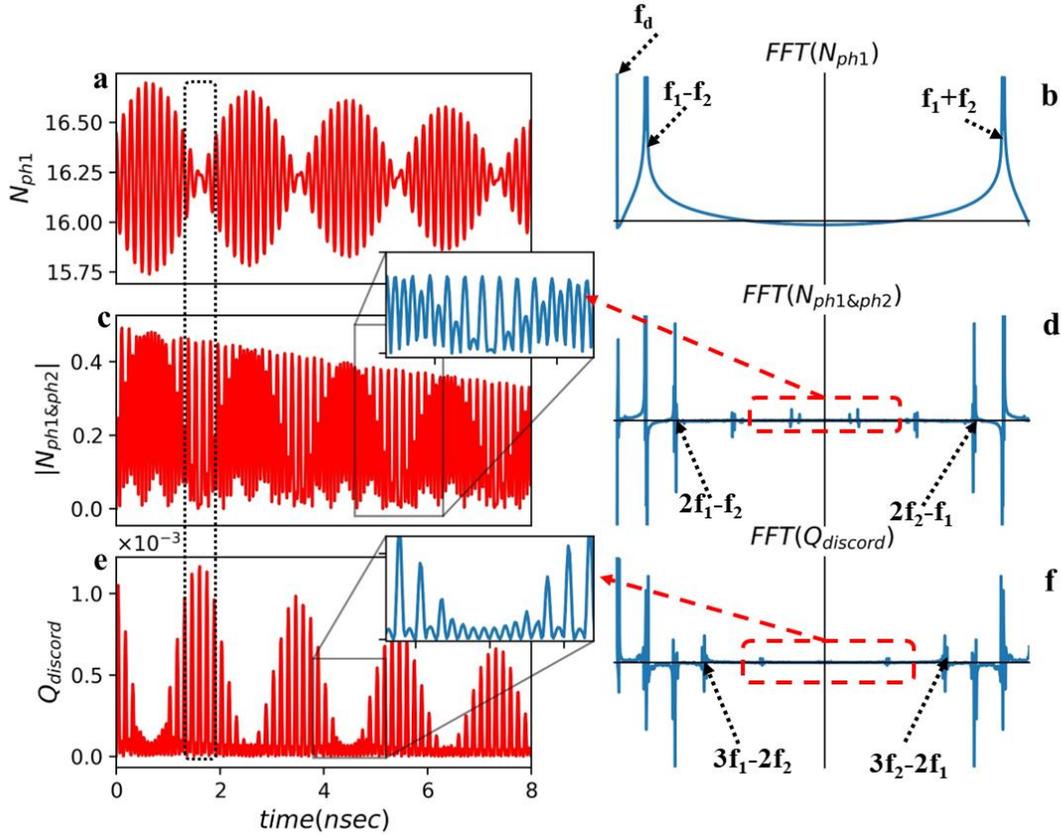

Fig. 3 a) and b) First LC oscillator average photon number and its FFT, c) and d) Phase-sensitive cross-correlation of the circuit and its FFT, e) and f) time evolution of quantum discord and the related FFT; the first and second oscillator-relating frequencies are $f_1$ = 5.57 GHz, $f_2$ = 5.77 GHz.

The other harmonics generated and shown in Fig. 3d, and Fig. 3f are $2f_1$-$f_2$, $2f_2$-$f_1$, $3f_1$-$2f_2$, and $3f_2$-$2f_1$. The results show that the quantum correlation is also created in the 3$^{rd}$ and 5$^{th}$ productions of the frequencies. It should be noted that the 3$^{rd}$ frequencies are so close to the system's original frequencies and cannot be easily filtered. In addition, other harmonics arise, illustrated with the dashed rectangle on the graphs. These harmonics are the properties of any real system, and they are created for different reasons such as re-mixing in the system and other effects. The other interesting thing in this graph is indicated by the vertical dotted rectangle shown in Fig. 3a, Fig. 3c, and Fig. 3e. This rectangle focuses on the area, where it is shown that when the quantum correlation ($Q_{discord}$) between modes is enhanced. It shows that to improve the quantum discord, the phase-sensitive cross-correlation should be enhanced, while at the same time, the oscillator's average photon number should be minimized. This is contributed to the mixing effect.

In the present design, the phase-sensitive cross-correlation is mainly manipulated with $\gamma_{q1\varphi2}$ and $\gamma_{q1q2}$. The rates mentioned handle the coupling between LC oscillator coordinates as $Q_1Q_2$ and $Q_1\varphi_2$ by which the splitter-like ($a_1a_2^+ \pm a_2a_1^+$) and amplification-like ($a_1a_2 \pm a_2^+a_1^+$) interactions [30-31] are created. The mentioned interactions generate 2nd harmonics as $f_1 \pm f_2$, the original harmonics that the quantum system (LC$_1$ coupling LC$_2$ through InP HEMT) operates with as a whole. The quantum correlation between modes mainly generates at these frequencies.

Suppose one designs a very narrow bandpass filter (or generally a low-pass filter) with a center frequency of around $f_1-f_2$. In that case, other frequencies' effects in the system will be subsided. This means that the quantum system designed needs no external mixer to down-convert the signals to be prepared for digitalization. Herein, using a very narrow bandpass filter, the only output signal at which the quantum correlation occurs is around 200 MHz. However, it is possible to engineer the circuit coupled to InP HEMT in such a way to attain $f_1-f_2 \sim 20$ MHz as a zero-IF band (zero-intermediate frequency). That is a suitable frequency band for sampling with a rate of around 100 MHz using an 8-bit ADC (analog-to-digital converter) [22]. As a result, we design a cryogenic quantum system able to generate zero-IF frequencies at which the quantum correlation is generated.

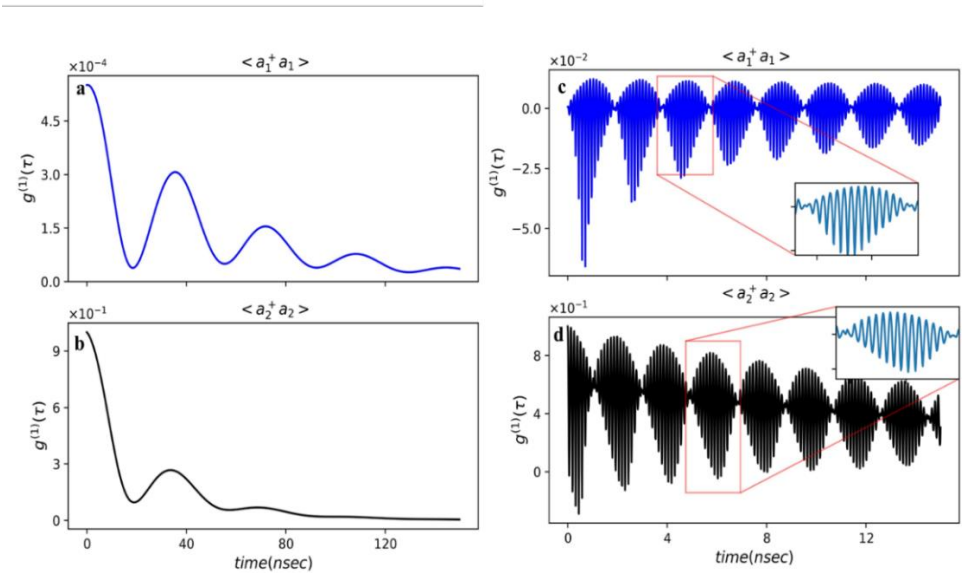

Fig. 4 The time evolution of the First-order coherent function a) and b) simple LC capacitively coupling to each other, c) and d) LC oscillators coupling through InP HEMT; up row for <$a_1^+a_1$> and down row for <$a_2^+a_2$>.

In the next part, the effects of the environment on the quantum system are studied. For this, the first-order coherent function is analyzed for the expectation value of the first and second oscillators. For instance, this quantifier shows that the value of the second oscillator initiated with a coherent state, how and by what rate decay to the thermal state due to the system's interaction with the bath. For a coherent state, $|g^1(\tau)| = 1$, while it is zero for an absolute thermal state. The simulation results are depicted in Fig. 4, in which the left column's graphs are for two simple LC oscillators capacitively coupling to each other, while the right ones are for the quantum system designed for this study. As mentioned, the first oscillator's initial state is put in the Fock state to compare the results with the second oscillator initiated with a coherent state. An apparent difference between the left and right graphs is that the mixing effect occurred due to the splitter-like and amplification-like interaction arising from the InP HEMT nonlinearity. This point was discussed in the latter section. Fig. 4b and Fig. 4d relate to the second oscillators initiated with the coherent state. Each figure shows that after a time, the amount of the first-order coherent function gradually decays due to the quantum system's interaction with the environment. Fig. 4c clearly shows that after 80 nsec, the second oscillator state decays completely to the thermal state. The same thing happened for Fig. 4d, while for a clear demonstration, the x-axis is limited to around 15 nsec.

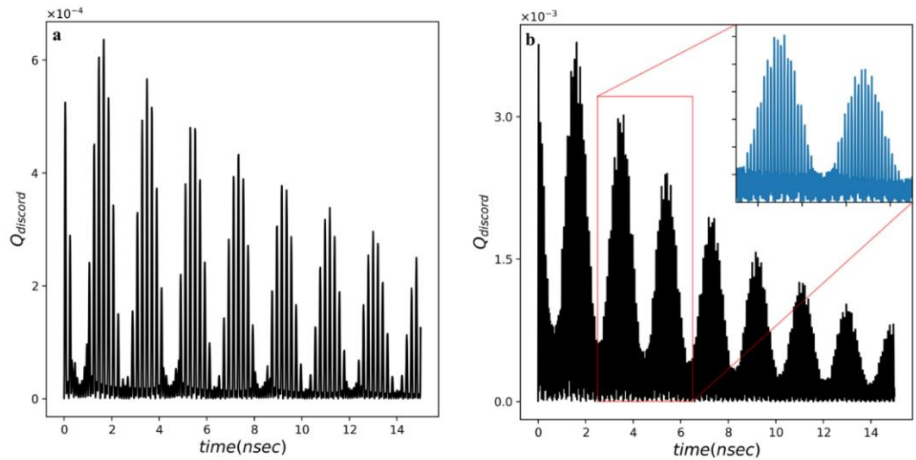

Fig. 5 Time evolution of the quantum discord for different values of the InP HEMT nonlinearity, a) $g_{m3} =$ 10 mA/V$^3$, b) $g_{m3} = 120$ mA/V$^3$.

Finally, the effect of the InP HEMT nonlinearity ($g_{m3}$) is studied, and the simulation results are shown in Fig. 5. The simulations are done for two different values of $g_{m3} = 10$ mA/V$^3$ and $g_{m3} = 120$ mA/V$^3$. An obvious point relating to Fig. 5a and Fig. 5b is that increasing the nonlinearity of InP HEMT leads to an increase in the quantum correlation between generated modes. In addition, it is shown in Fig. 5 that increasing $g_{m3}$ causes to grow in the frequency resonance of the oscillators. Of course, it has been theoretically demonstrated that $g_{m3}$ directly impacts the second oscillator frequency and impedance [21].

**Conclusions:**
One of the critical issues in quantum computing is the interface electronics between quantum processors operating at 10 mK and digital signal processing operating at room temperature. The interface electronics include the readout of the qubits and their controlling. HEMT is one of the trending technologies that has been used in the mentioned interface. For this reason, studying this important component and its quantum properties became interesting. This study mainly focused on the time evolution of the circuit's quantum correlation generated between modes that applies InP HEMT as a critical component. We initially derived the total Hamiltonian for the circuit, and using the Lindblad master equation, the time evolution of the related state for the quantum system was derived. In this study, all quantum mechanics simulations were done using Qutip toolbox in Python. The results showed that InP HEMT nonlinearity could operate as a mixer by which the coupled oscillator frequencies mix. In addition, we found that the quantum correlation could be established at 2$^{nd}$ and 3$^{rd}$ frequency of production. The second-order frequency production was the primary frequency that one could focus on and filter others to prepare the zero-IF frequency at which the quantum correlation occurred.

**References:**

1. B. Patra, R. M. Incandela, J. P. G. van Dijk, H. A. R. Homulle, L. Song, "Cryo-CMOS Circuits and Systems for Quantum Computing Applications," in IEEE Journal of Solid-State Circuits, vol. 53, no. 1, pp. 309-321, Jan. 2018, doi: 10.1109/JSSC.2017.2737549.
2. U. Kleine, J. Bieger and H. Seifert, "A low noise CMOS preamplifier operating at 4.2 Kelvin", Proc. 19th Eur. Solid-State Circuits Conf. (ESSCIRC), vol. 1, pp. 134-137, Sep. 1993.

3. E. Jeffrey, D. Sank, J. Y. Mutus, T. C. White, J. Kelly, R. Barends, Y. Chen, Z. Chen, B. Chiaro, A. Dunsworth,"Fast scalable state measurement with superconducting qubits", Phys. Rev. Lett., vol. 112, no. 19, pp. 190504, 2014.
4. A. Ferraro, L. Aolita, D. Cavalcanti, F. M. Cucchietti, and A. Acín, Phys. Rev. A 81, 052318-26, (2010).
5. A. Brodutch and D. R. Terno, Phys. Rev. A 81, 062103-6, (2010).
6. S.Pirandola, G. Spedalieri, S. L. Braunstein, N. J. Cerf, and S. Lloyd, Optimality of Gaussian Discord, Phys. Rev. Lett. 113, 140405-5, 2014.
7. G. Adesso and A. Datta, Quantum versus Classical Correlations in Gaussian States, Phys. Rev. Lett. 105, 030501-4, 2010.
8. P. Giorda and M. G. A. Paris, Gaussian Quantum Discord, Phys. Rev. Lett. 105, 020503-4, 2010.
9. Z. Merali, Quantum computing: The power of discord. Nature 474, 24–26 (2011). https://doi.org/10.1038/474024a
10. M. Zwolak, W. Zurek, Complementarity of quantum discord and classically accessible information. Sci Rep 3, 1729 (2013). https://doi.org/10.1038/srep01729.
11. U. Khalid, J. Rehman, H. Shin, Measurement-Based Quantum Correlations for Quantum Information Processing. Sci Rep 10, 2443 (2020). https://doi.org/10.1038/s41598-020-59220-y.
12. B. Dakić, Y. Ole Lipp, X. Ma, M. Ringbauer, S. Kropatschek, S. Barz, T. Paterek, V. Vedral, A. Zeilinger, Č. Brukner & Ph. Walther, Quantum discord as resource for remote state preparation, Nature Phys 8, 666–670 (2012). https://doi.org/10.1038/nphys237.
13. S. Barzanjeh, S. Guha, Ch. Weedbrook, D. Vitali, J. H. Shapiro, and S. Pirandola, "Microwave Quantum Illumination", Phys. Rev. Lett. vol. 114, pp. 080503-8, 2015.
14. A. Salmanogli, D. Gokcen, H.S. Gecim, Entanglement Sustainability in Quantum Radar, IEEE J. Sel. Top. Quantum Electron 26 (6), 1-11.
15. A. Salmanogli, D. Gokcen, Entanglement Sustainability Improvement Using Optoelectronic Converter in Quantum Radar (Interferometric Object-Sensing), IEEE Sensors Journal 21 (7), 9054-9062.
16. M. O. Scully, M. S. Zubairy, "Quantum Optics", Cambridge University Press, UK, 1997.
17. A. Salmanogli, D. Gokcen," Design of quantum sensor to duplicate European Robins navigational system", Sensors and Actuators A: Physical 322, 112636.
18. E. Cha, N. Wadefalk, G. Moschetti, A. Pourkabirian, J. Stenarson and J. Grahn, "InP HEMTs for Sub-mW Cryogenic Low-Noise Amplifiers," in IEEE Electron Device Letters, vol. 41, no. 7, pp. 1005-1008, July 2020, doi: 10.1109/LED.2020.3000071.
19. E. Cha, N. Wadefalk, P. Nilsson, J. Schleeh, G. Moschetti, A. Pourkabirian, S. Tuzi, J. Grahn, "0.3–14 and 16–28 GHz Wide-Bandwidth Cryogenic MMIC Low-Noise Amplifiers," in IEEE Transactions on Microwave Theory and Techniques, vol. 66, no. 11, pp. 4860-4869, Nov. 2018, doi: 10.1109/TMTT.2018.2872566.
20. A. Salmanogli, Squeezed States Generation using Cryogenic InP HEMT Transistor Nonlinearity, arXiv preprint arXiv:2204.08291, 2022.
21. A. Salmanogli, Entanglement Generation using Transistor Nonlinearity in Low Noise Amplifier, Quantum Sci. Technol. 7, 045026-35, 2022.
22. S. Barzanjeh, S. Pirandola, D. Vitali, and J. M. Fink, "Microwave quantum illumination using a digital receiver", Science Advances, 2020: Vol. 6, no. 19, eabb0451, DOI: 10.1126/sciadv.abb0451.
23. A. Salmanogli, Quantum Correlation of Microwave Two-mode Squeezed State Generated by Nonlinearity of InP HEMT, arXiv preprint arXiv:2211.01620, 2022.
24. I. Angelov, M. Garcia and H. Zirath, "On the performance of different HEMT cryogenic mixers," 1997 IEEE MTT-S International Microwave Symposium Digest, Denver, CO, USA, 1997, pp. 853-856 vol.2, doi: 10.1109/MWSYM.1997.602933.
25. M. Schefer, U. Lott, W. Patrick, H. Meier and W. Bachtold, "Passive, coplanar V-band HEMT mixer," Conference Proceedings. 1997 International Conference on Indium Phosphide and Related Materials, Cape Cod, MA, USA, 1997, pp. 177-180, doi: 10.1109/ICIPRM.1997.600083.
26. A. A. Houck, J. A. Schreier, B. R. Johnson, J. M. Chow, Jens Koch, J. M. Gambetta, D. I. Schuster, L. Frunzio, M. H. Devoret, S. M. Girvin, and R. J. Schoelkopf, Phys. Rev. Lett. 101 (2008) 080502-4.

27. Wallraff, A., Schuster, D., Blais, A. et al. Strong coupling of a single photon to a superconducting qubit using circuit quantum electrodynamics. Nature 431 (2004) 162–167. https://doi.org/10.1038/nature02851
28. J. R. Johansson, P. D. Nation, and F. Nori: "QuTiP 2: A Python framework for the dynamics of open quantum systems.", Comp. Phys. Comm. 184, 1234 (2013) [DOI: 10.1016/j.cpc.2012.11.019].
29. A. Salmanogli, Qubit Coupling to Reservoir Modes: Engineering the Circuitry to Enhance the Coherence Time, arXiv preprint arXiv:2205.13361, 2022.
30. J. Bourassa, F. Beaudoin, Jay M. Gambetta, and A. Blais, "Josephson-junction-embedded transmission-line resonators: From Kerr medium to in-line transmon," Phys. Rev. A vol. 86, pp. 013814, 2012.
31. A. Blais, R.Sh. Huang, A. Wallraff, S. M. Girvin, and R. J. Schoelkopf, "Cavity quantum electrodynamics for superconducting electrical circuits: An architecture for quantum computation," Phys. Rev. A vol. 69, pp. 062320, 2004.
32. A. Salmanogli, Squeezed States Generation using Cryogenic InP HEMT Nonlinearity, Journal of Semiconductors, 2023.

**Appendix A:**

The constants related to the Eq. 1 of the main article is defined as:

$$\gamma_{q_1q_2} = \frac{1}{4\sqrt{Z_1Z_2}} \left\{ \frac{C_BC_CC_A - C_C^3}{C_M^4} \right\}, \quad \gamma_{q_1\varphi_2} = \sqrt{\frac{Z_2}{Z_1}} \left\{ \frac{3C_BC_C^2 g_m - 2g_m C_B^2 C_A}{2C_M^4} \right\}, \quad \gamma_{\varphi_1} = \overline{I_{gs}^2} \sqrt{\frac{Z_1}{2\hbar}}$$

$$\gamma_{q_2\varphi_2} = \left\{ \frac{C_C^3 g_m + C_BC_CC_{A'}g_m - C_BC_C g_m C_A}{2C_M^4} \right\}, \quad \gamma_{q_1} = \left\{ \frac{C_B^2 C_{in}C_A V_{RF} - C_C^2 C_{in} C_B V_{RF}}{C_M^4} \right\} \sqrt{\frac{1}{2Z_1\hbar}}$$

$$\gamma_{\varphi_2} = \left\{ \frac{g_m C_C^2 C_{in} C_B V_{RF} - g_m C_B^2 C_{in} C_A V_{RF}}{C_M^4} - \overline{I_{ds}^2} \right\} \sqrt{\frac{Z_2}{2\hbar}}, \quad \gamma_{q_2} = \left\{ \frac{0.5 C_B C_C C_{in} C_{A'} V_{RF} + C_B C_C C_{in} C_A V_{RF} - C_C^3 C_{in} V_{RF}}{C_M^4} \right\} \sqrt{\frac{1}{2Z_2\hbar}}$$

$$\gamma_{q_1q_1\varphi_2} = \frac{g_{N2}}{C_M^4} C_B^2 \sqrt{\frac{\hbar Z_2}{2}} \frac{1}{2Z_1}, \quad \gamma_{q_2q_2\varphi_2} = \frac{g_{N2}}{C_M^4} C_c^2 \sqrt{\frac{\hbar Z_2}{2}} \frac{1}{2Z_2}, \quad \gamma_{q_2\varphi_2\varphi_2} = -\frac{g_{N2}}{C_M^4} 2g_m C_B C_c \sqrt{\frac{\hbar}{2Z_2}} \frac{Z_2}{2}$$

$$\gamma_{q_1\varphi_2\varphi_2} = \frac{g_{N2}}{C_M^4} 2g_m C_B^2 \sqrt{\frac{\hbar}{2Z_1}} \frac{Z_2}{2}, \quad \gamma_{\varphi_2\varphi_2\varphi_2} = \frac{g_{N2}}{C_M^4} g_m^2 C_B^2 \sqrt{\frac{\hbar Z_2}{2}} \frac{1}{2Z_2},, \gamma_{q_1\varphi_1\varphi_2} = \frac{g_{N2}}{C_M^4} 2g_m C_B C_c \sqrt{\frac{\hbar}{2Z_1}} \frac{\sqrt{Z_1 Z_2}}{2}$$

(A.1)

where $g_{N2} = g_{m2} + 6g_{m3}[\partial\varphi_1/\partial t]_{DC}$, $C_M^2 = C_B(C_A+C_N)-C_c^2$, $Z_1 = \sqrt{(L_1/C_{q1})}$, $Z_2 = \sqrt{(L_{2'}/C_{q2})}$, $C_C = C_{gd}$, $C_B = C_2 + C_{gd}$, $C_A = C_{in}+C_1+C_{gs}+C_{gd}$, $C_{A'} = C_A + C_N$, $C_N = g_{m2}[\varphi_2]_{DC} + 6g_{m3}[\varphi_2]_{DC}*[\partial\varphi_1/\partial t]_{DC}$, $\bar{I}_{gs}^2 = I_g^2 - I_j^2$, $\bar{I}_{ds}^2 = i_{ds}^2 + I_d^2 + I_j^2$. The thermally generated noises by the resistors and the current source are defined as $\bar{I}_g^2 = 4K_BT/R_g$, $\bar{I}_d^2 = 4K_BT/R_d$, $\bar{I}_j^2 = 4K_BT/R_j$, $i_{ds}^2 = 4K_BT\gamma g_m$, and $\bar{I}_i^2 = 4K_BT/R_i$, where $K_B$, $T$, and $\gamma$ respectively are the Boltzmann constant and operational temperature [32].

$C_1$ and $C_2$ are the total capacitors generated due to the external oscillators coupling to the InP HEMT. The full information about the details of the circuit and the contributed constants can be found in [32]. For a clear and intuitive view about the quantities and their relations, a figure is directly taken from [32], and presented in this section.

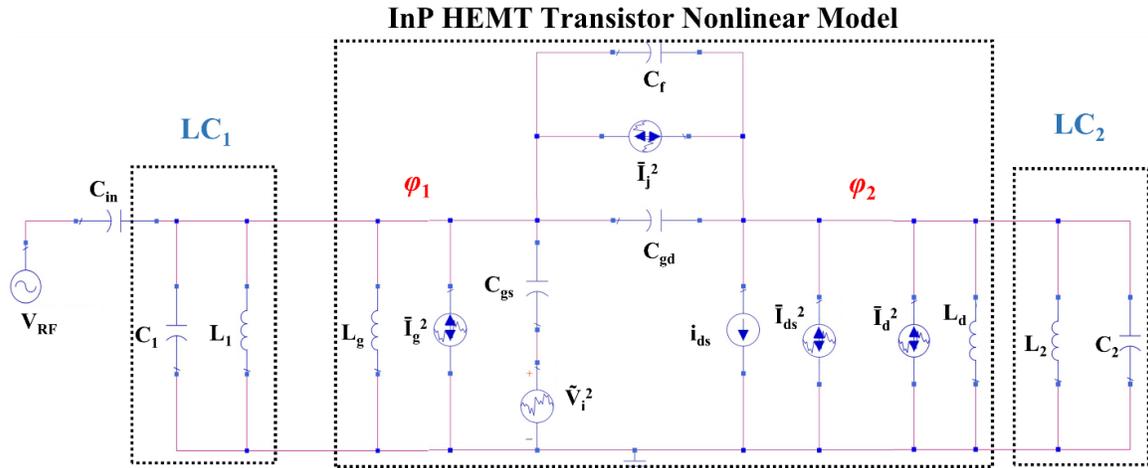

**Figure A1.** Complete system model; $L_{C1}$ coupling to $L_{C2}$ through InP HEMT Transistor operating at 5 K [32].